\documentclass[]{interact}
\usepackage{epstopdf}% To incorporate .eps illustrations using PDFLaTeX, etc.
\usepackage[numbers,sort&compress]{natbib}% Citation support using natbib.sty
\bibpunct[, ]{[}{]}{,}{n}{,}{,}% Citation support using natbib.sty
\usepackage{graphicx}
\usepackage[colorlinks=true,linkcolor=blue]{hyperref}
\usepackage{amsmath}
\usepackage{epsfig}
\usepackage{sidecap}
\newlength{\upit}\upit=0.1truein

\newcommand{\ltappr}{{{\lower4pt\hbox{$<$} } \atop \widetilde{ \ \ \ }}}
\newlength{\bxwidth}\bxwidth=1.5 truein

\newlength{\shift}
\newcommand{\fg}[3]
{
\begin{figure}[ht]
\vspace*{-0cm}
\includegraphics[width=\figwidth]{#1}
\vskip -0.3 cm
\caption{\label{#2}
#3
}
\end{figure}}

\newcommand{\fgh}[3]
{
\begin{figure}[h]
\centering
\vskip 0.0cm
\label{#2}
\includegraphics[width=\figwidth]{#1}
\vskip -0.2cm
\caption{
#3
}
\end{figure}
}
\newlength{\newwidthx}
\newcommand{\fgjpgmed}[3]
{\begin{figure}
\begin{minipage}[l]{0.6\textwidth}
\includegraphics[width=\newwidthx]{#1}
\end{minipage}\hfill
\begin{minipage}[l]{0.4\textwidth}
\caption{#2} \label{#3}
\end{minipage}
\end{figure}
}
\newcommand \bea {\begin{eqnarray} }
\newcommand \eea {\end{eqnarray}}
\newcommand{\bk}{{\bf{k}}}

\newcommand{\urs}{URu$_{2}$Si$_{2}$}
\begin{document}

\articletype{SCES Summary Talk}

\title{Theory Perspective: SCES 2016. }

\author{
\name{Piers Coleman\textsuperscript{a,b}\thanks{CONTACT
P.Coleman. Email coleman@physics.rutgers.edu}}
\affil{
\textsuperscript{a}Center for Materials Theory, Department of Physics and Astronomy,
Rutgers University, 136 Frelinghuysen Rd., Piscataway, NJ 08854-8019, USA}
\affil{\textsuperscript{b} Department of Physics, Royal Holloway, University
of London, Egham, Surrey TW20 0EX, UK.}}
%\date{}
%\pacs{72.15.Qm, 73.23.-b, 73.63.Kv, 75.20.Hr}
\maketitle
\begin{abstract}
New discoveries
and developments in almost every area of correlated electron physics
were presented at SCES 2016. 
Here, I provide a personal perspective on
some of these developments, highlighting some new ideas in
computational physics, discussing the ``hidden order'' challenges of
cuprate and heavy electron superconductors, the mysterious bulk
excitations of the topological Kondo insulator SmB$_{6}$ and  new
progress in research on quantum spin ice,  iron based
superconductors and quantum criticality. 
\end{abstract}

%
%\vfill\eject 
\section{Introduction: A Deluge of discovery and mystery.}\label{}
% SCES 16 Perspective, Hangzhou, China
%
% Theme: a time of discovery with exceptional opportunity for new insights.
%
% Theorists Eye View
% Materials Experiment, 
% Concepts, both sharp and fuzzy
% Models, Tools and Calculation (my precious!).
%
% A deluge of discovery ond mystery.
% Materials
% Concepts
% Modes.
One of the great insights of the 20th century physics
is that matter can acquire wholly unexpected new properties we call
{\sl ``emergence''}, from the collective behavior of interacting
quantum particles that lie within the material.
The quest to discover, understand and harness materials with such
novel {\sl emergent} properties is the 21st century frontier of
{\sl ``Strongly Correlated Electron Systems''} (SCES). In May 2016, 
the International Conference on Strongly Correlated Electron Systems,
SCES 2016 convened for the 
first time in China, at Zhejiang University in historic Hangzhou.\\

SCES is a field of research that continues to surprise its most
optimistic participants, with a deluge of
discovery and new mysteries that confront us each year. 
Here I'd like to share with you some of the exciting
developments that caught my eye at this meeting, complimenting Joe
Thompson's experimentally focused 
summary by emphasizing a theoretical perspective. 
\begin{figure}[ht]
\vspace*{-0cm}
\[
\includegraphics[width=\figwidth]{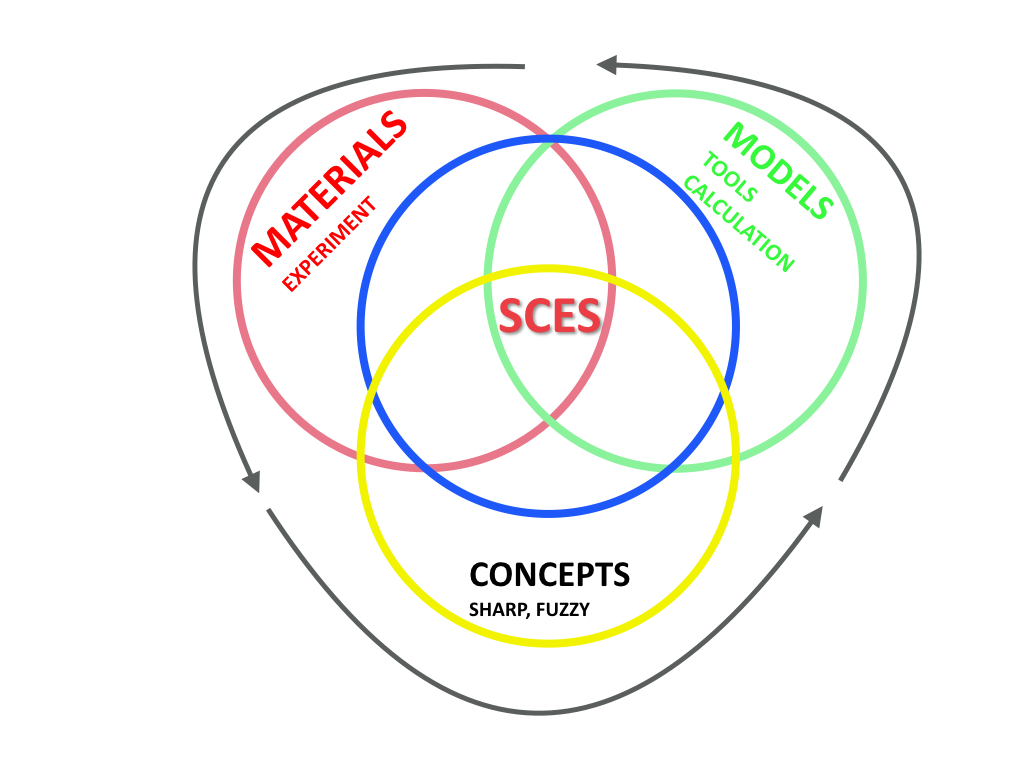}
\]
\vskip -0.2cm
\caption{\label{fig1}
The convective cycle of discovery in Strongly Correlated Electron Physics.}
\end{figure}

%\fg{fig1.jpg}{fig1}{}
%The convective cycle of discovery in Strongly Correlated Electron Physics.}

First a few words about the forces that drive discovery in our field.
Theoretical research into strongly correlated materials lies at the
intersection of three important areas: materials, concepts and
models (see Fig. \ref{fig1}).  Here are some of the topics that you might have heard about at
this meeting: 
\begin{itemize}
\item Materials and Experiment:  heavy electron, organic, cuprate, ruthenate,
iron-based, irridates and pyrocholores; laser ARPES, ultra-high magnetic
fields, dHvA, scanning tunneling microscopy, NMR, RIXS, Raman and
neutron spectroscopy to name a few.

\item Concepts: emergence, order parameter, topological order,
quasiparticle, Dirac, Majorana and Weyl Fermions, spin liquids and spin
ice, pseudo-gap, strange metals and quantum criticality, Mottness, Many Body
Localization, Skyrmion lattices and topological order. 

\item Models and methods: Heisenberg, Ising, Anderson, Hubbard,
Kondo and Kitaev models. Renormalization, Landau Ginzburg theory,
Path integrals, approximate $1/N$, $\epsilon$ and $1/D$ expansions,
Slave bosons and rotors, ab-initio density functional theory,  dynamical mean field
theory, Monte Carlo methods, density matrix renormalization and tensor
network methods.

\end{itemize}
One of the most powerful drivers of discovery is the convective
cycle between experiment and theory:
\begin{eqnarray}\label{l}
\hbox{new materials/experiment}&\Rightarrow&\hbox{wild ideas}\Rightarrow
\hbox{concepts}\Rightarrow \hbox{models}\Rightarrow
\hbox{predictions}\cr &\Rightarrow &\hbox{new materials/experiment}\dots 
\end{eqnarray}
Theorists often feel most secure working ``top-down'' from
well-established models
and {\sl ab-initio} methods, a vital part of the discovery process. 
Yet gems of discovery are 
to be found by those willing to
move outside their comfort zone, seeking new ideas and
insights in material discovery and experiments. To prospect 
amidst the real-world complexity of materials may sometimes seem
daunting, but it is here that the rewarding task of building models
and launching new interpretations often takes place.  
Here, the key tools are more phenomenological, such
as the Landau theory and scaling approaches.  The ``bottom-up''
approach 
theoretical condensed matter physics is 
valuable to experimentalists for it leverages their new data and 
occasionally provides the key to new concepts 
and the next generation of microscopic theories. 

Another aspect of discovery are interdisciplinary ideas,
borrowed from one another branch of physics or another class of
material and boldly
applied in a new context.  Such an approach can 
seed amazing breakthroughs.  For example, it is our fields'
adoption of Feynman diagram and field theory
methods born in Quantum Electrodynamics that drove the early
revolution in Many Body Physics\cite{Coleman2003}, including the BCS theory 
of superconductivity; aspirations towards similar breakthroughs
continue with efforts to use holographic methods borrowed from string
theory\cite{hartnoll}.  But even within condensed matter physics, 
who would have imagined that Kondo
insulators and the Quantum Hall effect would eventually become linked?
\begin{eqnarray}\label{l}
\hbox{Quantum Hall Effect}&\rightarrow &\hbox{Topological Matter}
\rightarrow \hbox{Topological Insulators}\cr
&\rightarrow& \hbox{Topological Kondo insulators},
\end{eqnarray}
or that the discovery of cuprate superconductors would inspire a 
plethora of new magnetic and superconducting materials? 
\begin{equation}\label{}
\hbox{cuprate superconductors}\rightarrow
\left\{\begin{array}{l}
\hbox{ruthenates}\cr
\hbox{115 heavy fermion superconductors}\cr
\hbox{irridates}
\end{array},
 \right.\end{equation}
or that the extension of the concept of vortex lattices in
superconductors into the magnetic domain would lead to skyrmion
lattices, a new development with serious applications? 
Each of these are important examples of the power of analogy and 
interdisciplinary collaboration that we need to strongly encourage in
SCES.

% 
% Interdis
\section{New Computational Methods}\label{}

In the 1960s, a combination of Monte-Carlo methods 
with Onsager's  exact solution to the Ising model,
paved the way for a grand revolution in understanding 
in statistical mechanics and critical phenomena\cite{domb}. 
One of our dreams today is to forge a similar breakthrough
in our study and simulation of interacting quantum systems. At this
meeting, two new developments caught my eye. 

\begin{itemize}
\item{\sl New progress on the fermion sign Problem of Quantum Monte
Carlo}.
In quantum systems, the Monte-Carlo method is a way of
computing thermodynamic partition function by writing it as a Feynman
sum over configurations weighted by the action of each configuration:
\begin{equation}\label{}
Z_{Q}= {\rm Tr}[e^{-\beta  H}] = \sum_{\{M\}} e^{-S[M]}.
\end{equation}
where the weight factor, $e^{-S[M]}= W_{\uparrow}[M]W_{\downarrow
}[M]$ is determined by the partition function $W_{\sigma
}[M]$ ($\sigma = \uparrow, \downarrow $) of fermions moving in the field of the order parameter 
$\{M \}$.  When fermions exchange, the wavefunction changes sign
a minus sign so the
quantum mechanical weight associated with a given configuration is
often negative.  Monte Carlo is only designed to deal with positive
probabilities, and if negative weight configurations proliferate, Monte Carlo
method breaks down. This is  the ``fermion sign problem''.  

At this meeting, Li, Jiang and Yao\cite{jiangX,jiang} injected some 
new optimism into this problem, pointing out that the fermion sign problem
is really basis dependent, which motivated them to search for new ways
to formulate the path integral to mitigate the
Fermion sign problem.  
One well known situation where sign problems disappear is the attractive 
Hubbard model, for which $W_{\uparrow}[M]W_{\downarrow }[M]>0$. In
this case, the Hubbard-Stratonovich Hamiltonian describing the fermion motion has
time-reversal symmetry. This guarantees that $W_{\downarrow
}[M]=W_{\uparrow}[M]^{*}$, so the resulting weight factor  $e^{-S[M]}=
|W_{\uparrow }[M]|^{2}>0$ is positive. 
(Unfortunately, this is not true for the repulsive $U$ Hubbard
model, because the Hubbard Stratonovich transformed Hamiltonian breaks
time-reversal symmetry.)
Yao et al. \cite{jiangX,jiang} have found that
that by splitting the fermion field into its real Majorana components,
$c = (\gamma_{1}+i\gamma_{2})$, one can classify and
identify new classes of
``Majorana Time-reversal Symmetries'' which guarantee that the sign
problem is absent.  I think this is a very interesting new direction,
and I suspect that we have more to learn by examining the
topological symmetries of models written in the Majorana language.

\item 
One area where there is scope for progress, is in the
implementation of Wilson's renormalization
concept into numerical methods.   With the development of density
matrix and tensor-network approaches, this is rapidly becoming a boom
area of computational condensed matter physics. 
One  advance that caught my eye was in
a simple implementation of renormalization in
the 
dynamical mean field approximation (DMFT). DMFT treats quantum many body
problems as an impurity, or cluster immersed n a
self-consistently determined bath.  At the heart of this approach, is
an impurity solver which computes electron self-energies. One popular
approach here is to evaluate impurity electron self-energies by
continuous-time Monte-Carlo methods. 
The random sampling of impurity histories
necessarily involves a lot of
virtual high energy processes that can, in principle be eliminated through the
use of the renormalization group.

At SCES2016, Changming Yue, Yillin Wang and Xi Dai\cite{yue} showed how 
an efficient continuous time solver for Dynamical mean theory can be
accomplished by integrating out high energy virtual charge
fluctuations, replacing them by an effective spin scattering
amplitude. For example, in an Anderson model, the virtual charge
fluctuations
\begin{equation}\label{}
f^{1}\rightleftharpoons f^{0}\rightleftharpoons f^{1}
\end{equation}
\fgh{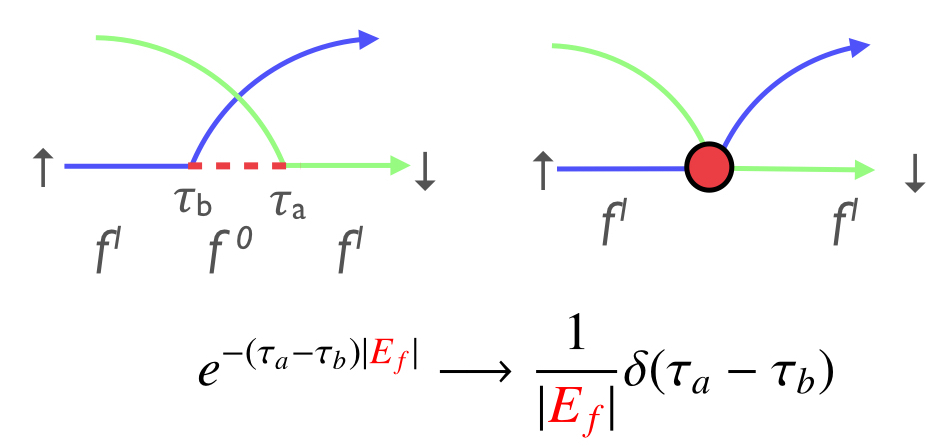}{fig2}{Replacement of a high frequency virtual charge
fluctuation by a single vertex enables the DMFT approach to
incorporate the key features of the Schrieffer-Wolff transformation.}

\noindent occurring at times $\tau_{b}$ and $\tau_{a}$ 
is associated with an amplitude $e^{- (\tau_{a}-\tau_{b})|E_{f}|}
$
where $|E_{f}|$ is the ionization energy of the atom. 
 When integrated
over time, this amplitude becomes $1/|E_{f}|
$, allowing the
replacement of the two virtual fluctuations by an effective spin
exchange vertex (Fig. \ref{fig2}).
%\begin{equation}\label{}
%e^{- (\tau_{a}-\tau_{b})|E_{f}|}\longrightarrow  1/|E_{f}|\delta (\tau_{a}-\tau%_{b})
%\end{equation}
This replacement is of course, the well-known ``Schrieffer Wolff
transformation''\cite{swolf} that folds the Anderson model into the Kondo model.
I was delighted to see that this same piece of physics has been made
to accelerate a DMFT solver, so that in effect, rather than asking  the
computer to carry out the Schrieffer Wolff transformation
millions of times a second, the renormalization is
included at the outset.  Yue et al demonstrated how their approach
could be used to efficiently model the low temperature 
ARPES spectral function of the Kondo lattice system
CeCoIn$_{5}$\cite{yue} down to much lower temperatures
This appears to be an important practical development and perhaps
first step along the road towards a fuller integration of  the DMFT with the 
Wilsonian renormalization group. 

\end{itemize}

\section{Competing and Hidden Order}\label{}

A continual challenge to SCES, is the discovery and
identification of new forms of quantum order.  Certain 
forms of order, such as
pair or multipolar density waves are very difficult to detect
directly, 
and may only be visible through their thermodynamic or indirect
influence on other excitations, giving rise to 
``hidden order'' (HO). 
The under-doped cuprates and URu$_{2}$Si$_{2}$ both exhibit such
hidden order. 
In the cuprate superconductors the development 
of a pseudo-gap\cite{pseudogap} in the density of states in the under-doped region of
the phase diagram is widely thought to be 
associated with one or more forms of hidden order (Fig. \ref{fig3x}).
In 
\urs, a  large moment antiferromagnet develops below 18K at 1.5GPa,
but when the pressure is released, the antiferromagnetic order is
replaced by hidden order, indicated by a large specific heat anomaly, 
with a Fermi surface geometry characteristic of a staggered order
parameter, yet without antiferromagnetism  (Fig. \ref{fig4x}) \cite{mydosh}.
Whereas pseudo-gap order in the
underdoped cuprates tends to suppress superconductivity, hidden
order in \urs\  actually induces it. 
In both cases, the nature of the hidden order and its 
relationship with superconductivity are a long-standing, and unresolved
mysteries.

At this meeting two new results on this topic caught my eye. 
\begin{itemize}

\figwidth=\textwidth
\fg{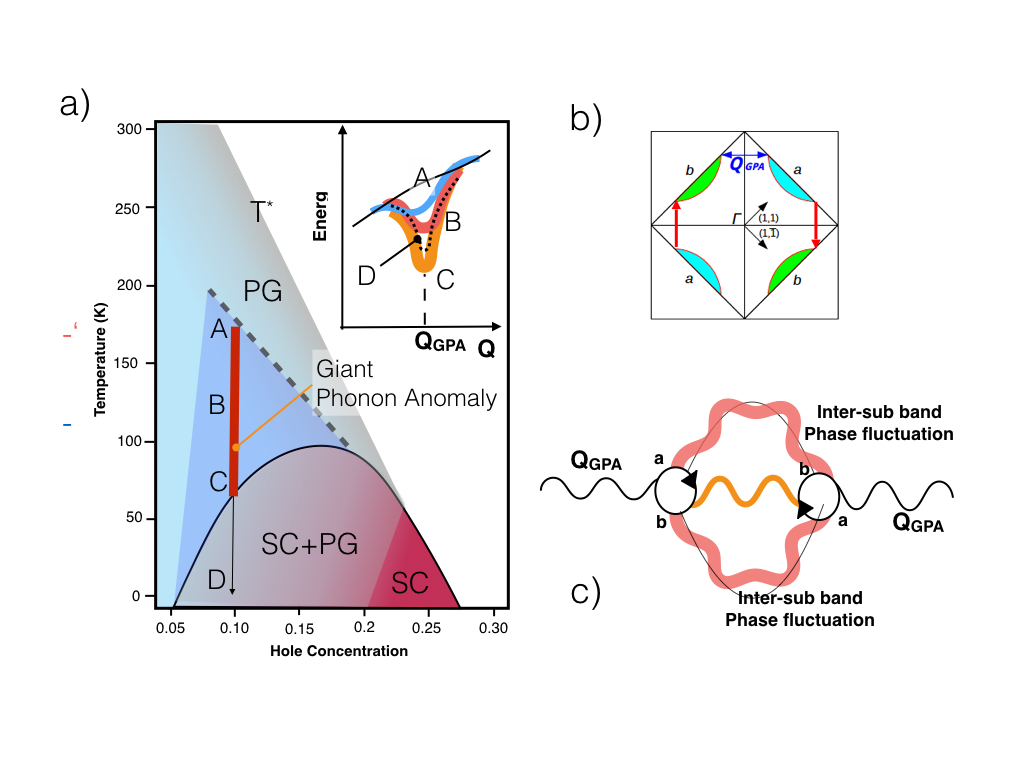}{fig3x}{(a) Phase diagram of under-doped cuprates,
sketched from \cite{vishik2012}, showing the pseudogap phase delineated by $T^{*}$.
The Giant Phonon Anomaly (GPA) develops at a lower temperature,
characterized by the growth of superconducting fluctuations. Inset:
phonons at wavevector $Q_{GPA}$ soften and broaden at onset $A$,
sketched from \cite{LeTacon:2013es}.
In the superconducting phase, the signal remains soft but sharpens.
(b) The wavevector $Q_{GPA}$ links the Fermi arcs causing inter-sub
band transitions. (c) Phase fluctuations (Leggett mode) couple to
inter-sub band fluctuations to produce the Giant Phonon anomaly
(sketched from \cite{Liu:2016ds})}

\item {\bf Giant Phonon Anomaly}. Maurice Rice discussed the under-doped cuprates, bringing to
light an unusual {\sl Giant Phonon Anomaly} (GPA) seen to develop
within the pseudo-gap phase\cite{LeTacon:2013es}. Below a certain temperature, considerably
below the pseudo-gap temperature ($T^{*}$), 
resonant X-ray scattering measurements show that phonons at particular
wavevectors $Q_{GPA}$ shift their energies and become damped. 
The onset of
the Giant Phonon Anomaly is seen to coincide with the observation of a
Josephson plasmon mode, associated with the development of interlayer pairing fluctuations\cite{PhysRevLett.106.047006}.
The wavevector $Q_{GPA}$ at which the anomaly develops matches with
the momentum required to scatter between the Fermi arcs of the
underdoped normal state (Fig. \ref{fig3x}a,b). 
Remarkably, the phonon broadening (but not
the frequency shift) disappears 
when the system goes superconducting. This has led 
Rice and collaborators\cite{Liu:2016ds} to propose a theory in which 
the phonon damping is driven 
by their coupling to an overdamped Leggett mode involving 
relative phase fluctuations between the developing
d-wave condensates formed in the (x,y)  and (x,-y)
directions (Fig. \ref{fig3x}c) .

\item {\bf Hidden order in \urs.} Girsh Blumberg's  group used Raman
scattering to shed new light on a possible 
relationship between Hidden Order to magnetism
in \urs. In both the low pressure HO phase and the high pressure AFM
phase, there is a collective Ising mode that can
be probed by both neutron and Raman Scattering.   By using iron doping
to tune from the HO to the AFM phase, they observe that the Ising mode
closes at the transition point between the two phases. 
This intriguing result suggests that

\fgjpgmed{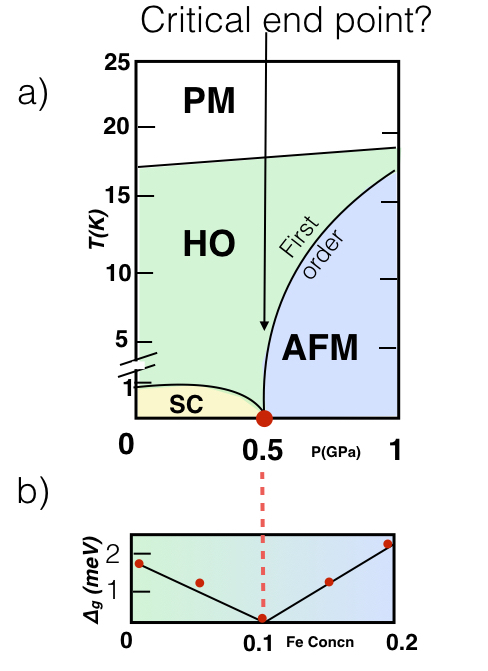}{(a) The pressure-temperature phase diagram of \urs, 
showing how the release of pressure causes antiferromagnetism to
transform into a state of Hidden order with superconductivity. (b) Sketch of the excitation
gap seen by Kung et al \cite{kung,kung2} 
using Raman scattering (using in iron-doping instead of
pressure), showing how the excitation gap collapses at the zero
temperature transition between Hidden order and AFM.}{fig4x}

\begin{enumerate}
\item the transition between the HO and the AFM at absolute zero is 
continuous, forming a critical end point at the bottom of the first
order line\cite{mucioetal} that separates the two phases at finite temperatures and 

\item at the transition, there is a gapless excitation  that links the 
the HO and the AFM phase, suggest that they 
can be continuously transformed into one-another.  If true, this has
many important implications - for example, it appears to suggest that 
the HO phase breaks time-reversal symmetry.
\end{enumerate}

\end{itemize}

These new developments pose many fascinating challenges to the
theoretical community. I am not, for example aware of any 
microscopic model for the interaction of
Raman modes with the magnetic and crystal field states of an
f-electron atom. Another aspect that
intrigues me, is whether the superconducting fluctuations in the
pseudo-gap  might also play a role in the strange metal phase that
develops at higher doping. Rice argued that high-momentum 
pair fluctuations (``pair density wave'') fluctuations play an
important role in the development of Fermi arcs. 
{\sl Could 
do  high-momentum  pair fluctuations 
persist outside the pseudogap region into the strange metal phase of
the cuprates? }
Such pair fluctuations tend to develop phase coherence between
electrons and holes moving in the same directions. Could they, I
wonder, be connected with the mysterious
observation of two transport relaxation rates, a longitudinal current
relaxation rate $\Gamma_{tr}\sim T$ which is linear, and a Hall
current relaxation rate $\Gamma_{H}\sim T^{2}$ which is 
quadratically dependent on temperature \cite{Coleman:1996eu} ? 

\section{SmB$_{6 }$ and Strange and Topological Insulators}\label{}

It is a tribute to the resiliency of the science of correlated quantum
materials, that a narrow gap insulator,  SmB$_{6}$  discovered almost
60 years ago,  is still generating new insights into quantum matter.

Over the past five years, it has become evident  that the strong
spin-orbit coupling in heavy fermion materials has the capacity to
drive topological phase transitions.  One
way in which this can take place, is through the band-crossing of odd
parity f-states and even parity d-states in Kondo insulators, giving
rise to a topological Kondo insulator\cite{dzero16}.  
In f-electron systems, there is the additional possibility
that interactions might drive qualitatively
new forms of topological quantum matter. At this conference SmB$_{6}$
seemed to offer just such a possibility.
\figwidth=\textwidth
\fg{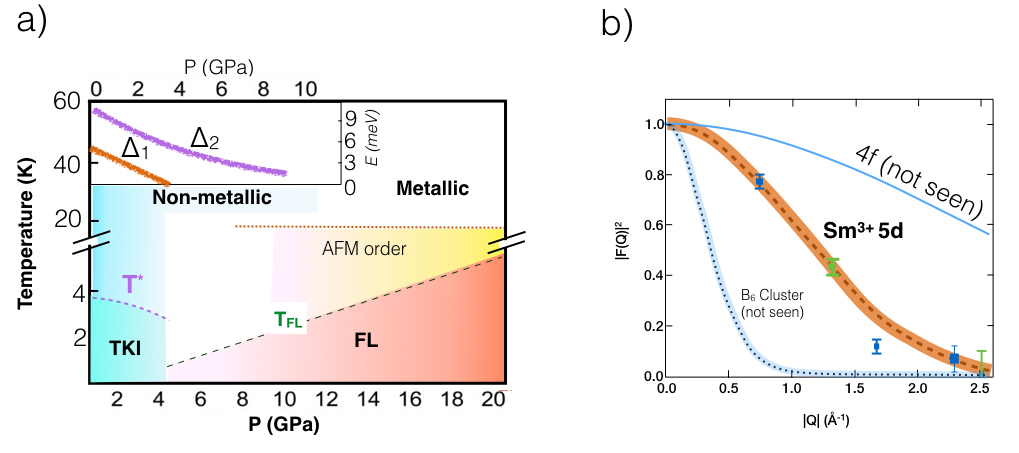}{fig5}{(a) Sketch of pressure phase diagram of Kondo
insulator SmB$_{6}$ measured in \cite{lilingsun,lilingsun2},
where T$^{*}$ is the onset temperature of the surface resistivity plateau,
T$_{FL}$ marks the onset of Fermi liquid behavior in the high pressure
metallic phase. $\Delta_{1}$ and
$\Delta_{2}$ are the transport and optical gaps, respectively. 
(b) Sketch of the 5d neutron form-factor seen in the SmB$_{6}$ magnetic
exciton from\cite{broholm}. Mysteriously, even though the exciton involves hybridized
f-electrons, the form factor is that of 5d electrons, as if the
f-electrons have a negligible g-factor. 
}

Here, I'd like to mention three separate developments: 
\begin{itemize}
\item The group of Liling Sun\cite{lilingsun} presented a new set of high-pressure
measurements that case new light into the 
phase diagram of SmB$_{6}$(Fig. \ref{fig5}a). This new
data confirms that the topological conductivity plateau surfaces up to
4GPa of pressure, where an insulator-metal transition occurs. 
Sun's group finds that the indirect transport gap $\Delta_{1}$ of the
insulator
collapses to zero at the insulator-metal transition, but the larger
hybridization (direct) gap $\Delta_{2}$ remains finite well into the
metallic regime. The characteristic Fermi scale of the metal appears
to grow linearly with pressure beyond the insulator-metal transition. 

\item Colin Broholm showed his groups measurement of a magnetic exciton
as the gap develops in SmB$_{6}$\cite{Alekseev:1993,Alekseev:1995fd,broholm}.  This is another indication of the
close vicinity of magnetic order.  A collective,
dispersing magnetic exciton
in SmB$_{6}$ was first observed by Alekseev and collaborators back in
the 1990s\cite{Alekseev:1993,Alekseev:1995fd}, but I think its fair to say that it is
only recently, that its  connection with a hybridized Kondo
insulator model has been understood in terms of a theoretical
model\cite{broholm,nikolic}. 
This data can be fit to a theory of an ingap
magnetic exciton formation between  hybridized d- and f-electrons\cite{broholm,nikolic}.
Yet mysteriously, in the neutron data reveals a
d-electron form factor, as if the vital f-electron part of the exciton
is entirely invisible to neutrons (Fig. \ref{fig5}b)  {\sl Is this because the f-electron's
have a very low g-factor, or is it an indication of something we have
overlooked in the bulk physics? }

\item Suchitra Sebastian and Lu Li presented the results of dHvA
measurements on this system. Lu Li argued that the angular dependence
of his  data imply two dimensional, topological surface states
consistent with the idea that SmB$_{6}$ is a strong topological
insulator. This alone is exciting. By contrast, Suchitra Sebastian argued that her group's results
suggest a gapless, albeit insulating bulk with 3 dimensional Fermi
surfaces. 
In support of this idea, 
Sebastian argued that the strong dependence of the effective
mass in the orbits at low temperatures is consistent with the large
bulk linear specific heat $C_{V}\sim \gamma T$ that has been long-observed in
this material and presented thermal conductivity data that indicate
that the thermal conductivity over temperature acquires a finite zero
temperature intercept $\left. 
{\kappa (T,H) /T}\neq 0\right\vert_{T\rightarrow 0}$ in a
magnetic field. 
\end{itemize}

These new developments suggest that while SmB$_{6}$ does
have robust topological surface states, there is something unusual
about the sub-gap bulk electrodynamics and magnetism of its
f-electrons.  Is it possible that the close vicinity to magnetism
transforms this material from a conventional band-gapped
topological insulator, into a {\sl strange insulator},
with a Fermi surface of gapless excitations that are responsible for
the linear specific heat, the thermal conductivity and dHvA
oscillations?  Were this true, it would  pose a most fascinating paradox:
for how can quasiparticles exhibit Landau quantization - a
semi-classical signature of the Lorentz force, while 
remaining an insulator?  My own view on this material is in a state of
evolution.  A year ago Erten, Ghaemi and I argued that the
observed properties of SmB$_{6}$ are consistent with topological
surface states, combined with a break-down of the Kondo effect at the
surface\cite{Erten:2016ia} but the latest data, with a far more detailed
study of the angular dependence of the low frequency modes, and the
unusual thermal conductivity  suggest that this
view needs re-evaluation.  The new results 
motivate considering the possibility of a new kind of 
insulator, one in which a neutral shadow of the original Fermi surface
remains unhybridized, yet insulating.  Could this be a gapless
spinon Fermi surface, perhaps a $Z_{2}$ spin liquid
lurking within the material?  Alternatively, the idea of a Majorana
Fermi surface, proposed 
by Eduardo Miranda, Alexei Tsvelik and myself many years ago and
renovated by Ganputhy Baskaran may prove
useful\cite{Tsvelik:1993us,Baskaran:2015mja}. According to this
radical view, SmB$_{6}$ would be a failed superconductor with neutral current
excitations that can link selectively to the Lorentz force,
thanks to an almost broken $U (1)$ gauge
symmetry\cite{inprogress}.  

I mention in passing that there may be many other heavy fermion
systems with topological surface states.  We didn't hear much about
these possibilities at this meeting, but plateau conductivities are
present in old data on Sb doped CeNiSn\cite{Slebarski} and Ce$_{3}$Pt$_{4}$Bi$_{3}$
under pressure\cite{Cooley_PRB97},
which may also be topological Kondo insulators, protected by their
non-symorphic crystal symmetries\cite{mobius} 

\section{Quantum Spin Ice}\label{}
\fg{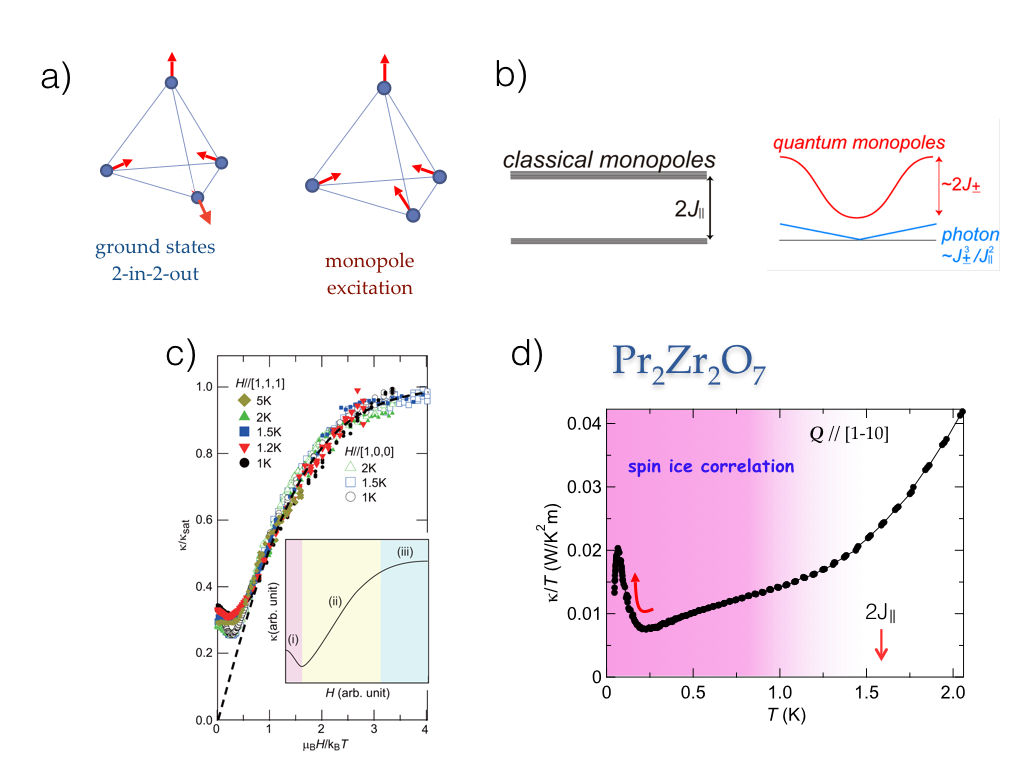}{fig6x}{a) In ``spin ice'' the ground-state of each
tetrahedron satisfies the ice rules (two spins in, two spins out)\cite{gingras14}.
``Monopole'' excitations correspond to three out/in, one in/out. b)
Classical monopoles are localized and have an excitation energy of
$2J_{\parallel}$.  Quantum monopoles are itinerant, and theory also
predicts the development of emergent ``photon'' excitations. c)
Field dependence of thermal conductivity in Yb$_{2}$Ti$_{2}$O$_{7}$
from \cite{tokiwa} at various temperatures. d) Temperature dependence
of thermal conductivity $\kappa/T$ in spin ice compound
Pr$_{2}$Zr$_{2}$O$_{7}$ reported by Tokiwa et al.,  showing
marked upturn below 0.2K which is interpreted as a signal of the spin-photon.
}

Frustrated spin systems offer the opportunity to 
observed strongly correlated behavior in quantum materials 
without the complications of charge motion. One of the areas of
particular interest in recent years has been that of ``spin ice''
pyrochlore magnets.  These uniquely frustrated magnets contain spins
arranged on the vertices of tetrahedra arranged in pyrochlore lattice,
which obey the ``two-in, two-out'' ice rules. Excitations out of the
spin-ice manifold form magnetic monopoles with an excitation energy of
$2J_{\parallel }$ (Fig. \ref{fig6x} a \& b).  This has raised the fascinating possibility that
when quantum fluctuations are included, spin ice will melt, to form a
quantum spin liquid. Such quantum spin ice is predicted to have
itinerant monopoles and an ``photon'' mode associated with its
emergent electric and magnetic fields\cite{PhysRevB.69.064404}. 

At this meeting I was particularly fascinated by new results on two spin-ice 
pyrochlore lattices Yb$_{2}$Ti$_{2}$O$_{7}$ and
Pr$_{2}$Zr$_{2}$O$_{7}$ 
by Yoshi Tokiwa and
collaborators \cite{tokiwa,nakatsuji2}.  
 Yb$_{2}$Ti$_{2}$O$_{7}$ ferromagnetically orders
below $T_{C}=0.2K$, but above this temperature, based on their
measurements, Tokiwa et al. propose 
it is pyrochlore quantum spin liquid with itinerant quantum
monopoles. By contrast, Pr$_{2}$Zr$_{2}$O$_{7}$ does not magnetically
order down to the lowest temperatures.

In Yb$_{2}$Ti$_{2}$O$_{7}$\cite{tokiwa} Tokiwa et al have measured the thermal conductivity $\kappa (T,H)$versus
field $H$ and temperature $T$.  At high temperatures, the thermal conductivity
is a function of $\mu_{B}H/T$, and can be understood as a result of
spin-phonon scattering. However, at low fields, the thermal
conductivity $\kappa (H)/T$ decreases with applied field, which the
authors interpret as a result of Monopole thermal conductivity
(\ref{fig6x} c ).
This is a very interesting result, and it will be interesting to
examine whether at still lower temperatures, a thermal conductivity
from the spin-photons can be observed.

In the sister comound Pr$_{2}$Zr$_{2}$O$_{7}$\cite{nakatsuji2}, Tokiwa reports
that he system displays no discernable long range order,
and that  the measured inelastic neutron scattering dominates  90\% of the scattered
intensity, consistent with a quantum spin ice compound. 
Perhaps most excitedly, the thermal conductivity $\kappa/T$ displays a marked
upturn at low temperatures (Fig. \ref{fig6x}d) that may be the first signs of the fabled
spin-photon excitation. There is a hope that future neutron
measurements will be able to resolve and confirm the presence of this
excitation. This is a developing story to keep your eyes on.

\section{Iron Based Superconductors}\label{}

Unlike the cuprate superconductors, where we have unambiguous evidence
for the symmetry and structure of the pair wavefunction, in the iron based
superconductors, this issue is still a matter of continuing
discussion and fascination\cite{Si:2016kl}.  Of particular interest, is the mechanism by which the
pair condensate overcomes the considerable Coulomb repulsion between
electrons on the iron sites. 

Most iron based superconductors are fully gapped, and the
canonical theory for their gap structure supposes that it has
s$^{\pm}$ structure, with an s-symmetry. In the s$^{\pm }$ scenario,
spin fluctuations exchanged between the electron and hole bands 
lead to a sign difference between the electron and
hole pockets. This sign difference gives rise to a suppression of the
local s-wave pair density, and thereby overcomes the strong onsite
Coulomb interactions. 

The discovery of a number of ``bad-actor'' iron-based superconductors
which only have electron, or hole bands, poses a difficulty to the 
s$^{\pm}$ pairing mechanism.   
One interesting idea in this respect, is the possibility that the
orbital quantum numbers of the paired quasiparticles can entangle with
the pair condensate,  in a fashion reminiscent of the spin
entanglement present in triplet superfluids or
superconductors\cite{Yin:2014fj,SiPRB2014,Ong:2016jf,Nourafkan:2015wa}.  In an
conventional scenario the pair condensate wavefunction is diagonal in
the orbital indices
\begin{equation}\label{}
\langle d _{\mu \uparrow} (\bk )d_{\nu\downarrow }\rangle \propto
\delta_{\mu\nu} \Delta (\bk )
\end{equation}
whereas in an orbitally entangled scenario, the 
pair wavefunction has a non-trivial orbital dependence
\begin{equation}\label{}
\langle d _{\mu \uparrow} (\bk )d_{\nu\downarrow }\rangle \propto
\sum_{\Gamma} \Delta_{\Gamma} (\bk ) \alpha^{\Gamma}_{\mu\nu},
\end{equation}
where  the $\alpha^{\Gamma}$ are matrices in orbital space and the
functions $\Delta_{\Gamma}$ are their corresponding gap functions.

\fg{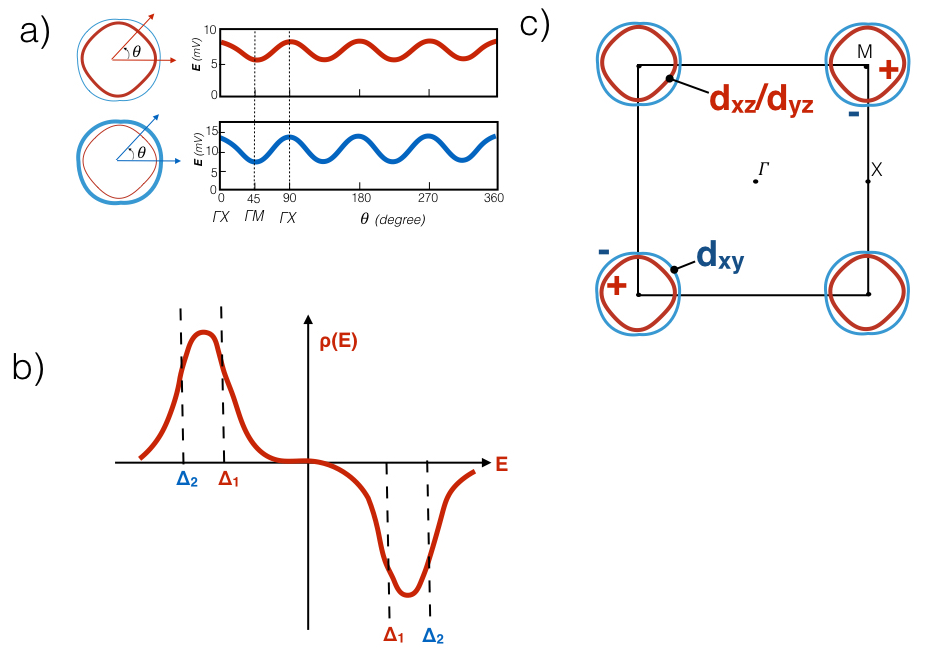}{fig7}{Sketch of STM Data taken on
(Li$_{1-x}$Fe$_{x}$)OHFeSe from \cite{hwen16a} a)showing two fully gapped
electron pockets around $M$ points b) sketch of antisymmetric
contribution to the quasiparticle-interference spectrum, peaked
between energies $|\Delta_{1}|$ and $|\Delta_{2}|$, a signature of
potential scattering between two Fermi surfaces with opposite gap sign c)
inferred antiphase structure of the gaps on the hole pockets.}

One of the iron-based superconductors with exclusively electron pockets, 
is the single layer FeSe iron chalcogenide, with a transition
temperature $T_{c}$ in excess of 60K, yet which only has electron 
pockets centered around the $M$ points.
At this meeting, the group of Hai Hu Wen announced a new set of
scanning tunneling spectroscopy measurements on the layered iron chalcogenide
(Li$_{1-x}$Fe$_{x}$)OHFeSe, a bulk analogue 
of FeSe with a $T_{c}$ of 40K that appears to
confirm an orbital dependence to the pair wavefunction.  
These measurements
(Fig. \ref{fig7} a) 
confirm that the confirm that the electron pockets 
pockets have a fully developed gap\cite{hwen16a,hwen16b}. The large
ratio $2\Delta /T_{c}\sim 9$ indicates that the pairing is in the
strong coupling limit. 
Wen's group employ some recent theoretical analyses of the effects of
pairing on interband quasiparticle scattering\cite{Hirschfeld:2015do}
to show that gaps on the two hole pockets are in {\sl antiphase}.
Here the basic idea is that interband scattering between two Fermi
surfaces with opposite pairing signs gives rise to a correction to the
density of states that is antisymmetric in energy and peaked at
energies $E\in [|\Delta_{1}|,|\Delta|_{2}]$.  It is this feature that
is observed in the data (Fig. \ref{fig7} b).
This is particularly interesting, for the outermost electron  pocket has a
predominantly $d_{xy}$ character, while the innermost hole pocket have
$d_{zx}/d_{zy}$ character. These results seem to be consistent
with orbital antiphase pairing.

These new results will doubtless stimulate further work in this
direction. Some of the questions that they pose are:
\begin{itemize}
\item Is there a non-trivial dependence of the pairing on the other
$d_{zx}$ and $d_{zy}$ orbitals? 
\item Does the strong-coupling limit of the pairing in the iron-based
superconductors have a local interpretation, e.g in terms of localized
bosons or perhaps composites of pairs and spins? 
\end{itemize}

\section{Strange Metals and Quantum Criticality}\label{}

Historically, quantum criticality in strongly correlated,
particularly heavy fermion systems has been cast in terms of the
Doniach competition between the RKKY and the Kondo interactions.
Over the past decade or so, our community has become increasingly
aware of a second quantum criticality axis - that of 
quantum and geometric frustration, which can also 
drive local moment systems from ordered magnetism
into paramagnetic spin liquids or valence bond solids. The combined
influence of these two effects is notionally mapped out in the
global phase
diagram\cite{senthil04,Si:2006dv,Lebanon:2007in,Coleman:2010cf} for
the Kondo lattice (Fig. \ref{fig8}).

SCES 2016 was marked by the entry of several new materials that
illustrate different aspects of the global phase diagram for
quantum criticality. In particular:
\begin{enumerate}
\item {\bf 1 D spin liquid behavior in a heavy fermion compound}. The observation  of spinons in the quasi-one dimensional
heavy fermion magnet Yb$_{2}$Pt$_{2}$Pb \cite{Wu:2016du}.  This material
illustrates the feasibility of spin-liquid formation within a strongly
spin-orbit coupled heavy fermion material (See: Fig. \ref{fig8} a ). 
\item {\bf Frustration tuned non-Fermi liquid behavior} was observed
in the heavy fermion Kagome-lattice compound CeRhSn
\cite{Tokiwa:2015kd}. Here, careful Gr\" uneisen parameter measurements
reveal that the application of strain which selectively relieves the
Kagome-lattice frustration, drives the strange metal state back into a
Fermi liquid (Fig. \ref{fig8} b). 

\item {\bf Strange metals, robust against pressure}. At this meeting,
two strange metallic Yb heavy fermion compounds, the hexagonal layered
system $\beta -$YbAlB$_{4}$\cite{Tomita:2015hla} and the heavy fermion
quasi-crystal YbAlAu\cite{Deguchi:2012hea} were discussed.  Both of these
systems exhibit field-tune quantum criticality - namely power-law behavior in their specific
heat and magnetic susceptibility, which is field tuned and reverts to
Fermi liquid behavior in the smallest applied magnetic
fields. Remarkably, the application of pressure fails to remove this
quantum critical behavior in either compound, suggesting the formation
of a strange metallic phase.  Also, Deguchi finds that 
strange metallic behavior is only present in pristine quasicrystals of
YbAlAu\cite{Deguchi:2012hea} while the approximant crystal
Au$_{51}$Al$_{35}$Yb$_{14}$, with precisely the same local structure,
but a finite unit cell size, is a Fermi liquid (Fig. \ref{fig8} c).
\end{enumerate}

This is clearly a gold-mine for future theoretical activity.  I would
like to mention a few key questions and observations taken away from
these results
\begin{itemize}
\item CeRhSn, $\beta -$YbAlB$_{4}$\cite{Tomita:2015hla} and YbAlAu\cite{Deguchi:2012hea} are each examples of
metals in which the non-Fermi liquid behavior
persists over a range of parameters. From these materials I think it
is clear that we should take the idea of strange metal phases
seriously. What is the theoretical basis for such phases? 
\item  What is the role of valence fluctuations? Kazumasa Miyake and
Shinji Watanabe\cite{watanabe1,watanabe2}
argue that quantum critical valence fluctuations are the driver 
for strange metallic behavior. Certainly, the
strange metallic behavior in 
$\beta -$YbAlB$_{4}$\cite{Tomita:2015hla} and YbAlAu is clearly separated from magnetic
quantum criticality. We need to understand if valence fluctuations can
account for strange metallic phases, or whether alternative mechanisms
are needed.

\item Why is the pristine quasicrystal YbAlAu a strange metal, yet its
approximant Au$_{51}$Al$_{35}$Yb$_{14}$\cite{Deguchi:2012hea} a Fermi liquid?  Since there is no local difference in the
electronic structure or chemistry of these two compounds, we might speculate
that the quasicrystal must contain some kind of electronic or
magnetic criticality that is cut-off by the finite size of the unit
cell in the approximant. 
\item We heard from Yifeng Yang\cite{Yang:2008dm} about the two-fluid phenomenology of
heavy fermion materials used to understand heavy fermion systems on
the brink of magnetism.  One of the most striking examples of such
behavior is CeRhIn$_{5}$ in which the Cerium f-moments homogeneously
time-share between local moment magnetism and superconductivity.  I
think the time has come to think of ways to try to cast these ideas
into the language of a wavefunction and I'd like to encourage the
community ways of doing this.  
My own favorite, formulated by my former student Aline Ramires, 
 is the possibility of a
Gutzwiller projected product of a bosonic magnetic fluid and a
fermionic Kondo liquid
\[
|\Psi\rangle = P_G \bigl(\vert \Psi_F\rangle \otimes |\Psi_B\rangle\bigr),
\]
where  $P_{G}$ is a Gutzwiller projection that entangles the 
product of a fermionic $\vert  \Psi_{F}\rangle $
and bosonic $\vert  \Psi_{B}\rangle $ representation of the spin
background\cite{aline}.
My main point however, is that the theory community needs 
to be thinking about new approaches and next 
generation wavefunctions that may help us to understand the
entangled nature of magnetism and superconductivity. 

\end{itemize}
\figwidth=1.05\textwidth
\fg{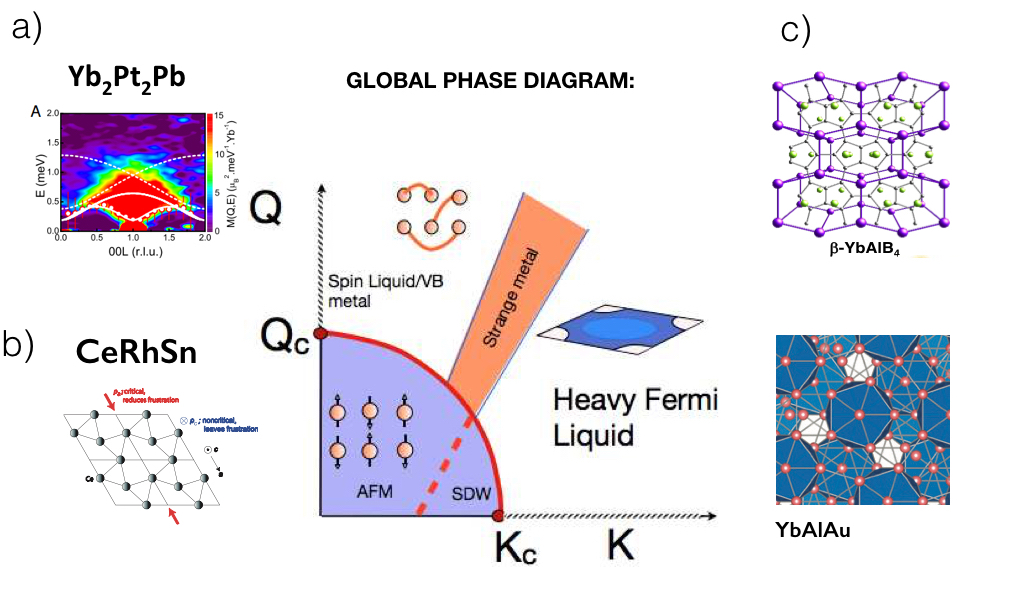}{fig8}{Global Phase diagram for heavy fermions
\cite{senthil04,Si:2006dv,Lebanon:2007in,Coleman:2010cf} with horizontal 
``Kondo'' and quantum frustration (Q) axes. a) Spinon excitations in
the inelastic neutron scattering spectrum of Yb$_{2}$Pt$_{2}$Pb,
figure\cite{Wu:2016du,Wu:2016duX} 
 b) CeRhSn
in which only strain that relieves the frustration is a relevant
perturbation to the non-Fermi liquid ground-state \cite{Tokiwa:2015kd}
 c)  the layered crystal
$\beta -$YbAlB$_{4}$\cite{Tomita:2015hla} (Figure courtesy of Yosuke Matsumoto) and the quasicrystal YbAlAu
\cite{Deguchi:2012hea} (Figure courtesy of Kazuhiko Deguchi) both exhibit
field-tuned quantum criticality, disconnected from magnetic
criticality and robust against pressure. }

%\section{Introduction}\label{} 

\noindent {\bf Acknowledgments}

This research was supported by the United States Department of Energy
Basic Energy Sciences grant DE-FG02-99ER45790 and
United States National Science Foundation grant NSF DMR 1309929.
This article was written while 
at the Aspen Center for Physics, which is supported by National Science Foundation grant PHY-1066293.
I should like to thank Meigan
Aronson, Colin Broholm, Girsh Blumberg, Po-Yao Chang, Xi Dai, Onur Erten, 
Gilbert Lonzarich, 
Suchitra Sebastian,
Liling Sun, Joe Thompson, Yoshi Tokiwa, Hai Hu Wen and Changming Yue
for discussions related to their work and this talk. My thanks to the
organizers of SCES 2016, which was truly a memorable conference. 
Finally, my sincere apologies to the
authors of the many wonderful talks and posters that I heard and
visited, but was unable to include in this perspective. Your work is great
and I look forward to learning how it now develops.

\bibliographystyle{tfq}

%\bibliography{scesbib}
\end{document}